%% file: rtwo_shil.tex
\documentclass[sigconf, pbalance=true]{acmart}
\usepackage{graphicx}
\usepackage{float}
\usepackage{booktabs, array}
\usepackage{microtype}
\usepackage{caption}

\setcitestyle{sort}

\usepackage{circuitikz}
\usepackage{tikz-3dplot}
\usepackage[detect-weight]{siunitx}
\usepackage{setspace}
\usepackage{multirow}
\usepackage{mathtools}

\usetikzlibrary{backgrounds}
\usetikzlibrary{arrows.meta}

\graphicspath{{figures/}}
\newcommand*{\tabref}[1]{\tablename~\ref{#1}}
\newcommand*{\figref}[1]{\figurename~\ref{#1}}
\newcommand*{\secref}[1]{Section~\ref{#1}}

\ccsdesc[500]{Hardware~Clock generation and timing}
\ccsdesc[500]{Hardware~Power and thermal analysis}
\ccsdesc[500]{Hardware~Emerging architectures}

\setcopyright{none}
\renewcommand\footnotetextcopyrightpermission[1]{%
  \footnotetext[0]{© Nicholas Sica 2026. This is the author's version
    of the work. It is posted here for your personal use. Not for
    redistribution. The definitive Version of Record was published in
    Great Lakes Symposium on VLSI (GLSVLSI) 2026, http://dx.doi.org/10.1145/3787109.3815289.}}

\begin{document}
\title{ROA-Based Subharmonic Injection Locking for Oscillator-Based
  Ising Machines}

\author{Nicholas Sica}
\orcid{0009-0000-9321-7038}
\affiliation{
  \institution{Drexel University}
  \city{Philadelphia}
  \state{Pennsylvania}
  \country{USA}
}
\email{njs82@drexel.edu}

\author{Baris Taskin}
\orcid{0000-0002-7631-5696}
\affiliation{
  \institution{Drexel University}
  \city{Philadelphia}
  \state{Pennsylvania}
  \country{USA}
}
\email{bt62@drexel.edu}

\begin{abstract}
  This paper introduces on-chip integrated rotary traveling wave
  oscillators~(RTWOs) organized into rotary oscillator array~(ROA)
  bricks as an external perturbation to induce subharmonic injection
  locking~(SHIL) in oscillator-based Ising machines~(OIMs). The
  implementation of SHILs on chip is challenging, as the frequency of
  SHILs must be multiples of the operating frequency of the OIM nodes,
  with on-chip variations affecting the phase, degrading the SHIL
  process. This impedes the scaling of OIM implementations, regardless
  of the topology of Ising nodes, coupling or graph mapping
  mechanisms. The ROA brick topology implementation of RTWOs generates
  high frequency signals that are shown to provide a stable
  \qty{2.31}{\GHz} SHIL signal under process, voltage, and
  temperature~(PVT) variations. Under PVT variations, distributed ring
  oscillator-based SHILs~(ROSC-SHIL) fail to perform injection locking
  while the proposed ROA brick-based SHIL~(ROA-SHIL) preserve 93\% to
  97\% accuracy~(the same accuracy of an ideal SHIL signal) in the OIM
  solutions of a sample 324-node max-cut problem. The driving strength
  and floorplan of the ROA brick are also shown to be amenable for
  scaling with an energy-to-solution impact of \qty{2.49}{\nano\joule}
  for the proposed ROA-SHIL.

\end{abstract}

\keywords{Ising machine, combinatorial optimization, resonant
  clocking}

\maketitle

\section{Introduction}
Ising machines~\cite{ernst_ising} are promising for speed and energy
efficiency in solving combinatorial optimization
problems. Oscillator-based Ising
machines~(OIMs)~\cite{oim_roychowdhury} use phases of CMOS oscillator
nodes as spins of an Ising machine, coupled together through resistive
links to decrease energy in a system. The settled groups solve the
combinatorial optimization problem. In absence of an external
perturbation of subharmonic injection
locking~(SHIL)~\cite{oim_roychowdhury}, the oscillators in Ising
machines settle to a continuum of phases instead of discrete
groups. SHIL injects a predetermined signal operating at a subharmonic
of the oscillator frequency into the oscillator nodes. The discrete
values to which the oscillator nodes settle with the external
perturbation of the SHIL signal correspond to Ising spin values, $s_i$
= 1 or -1. Recent CMOS implementations of OIMs~\cite{kim_1968_nodes,
  ising_image_recognition} utilize an external~(i.e.\ off-chip) signal
to induce SHIL on ring oscillator-based Ising nodes. While an external
signal is useful in demonstrating the utility of Ising machines at
small scales, on-chip integration of SHIL generation circuits and
distribution are essential for scalability. Previous research has
shown the effect of methodically changing the amplitude of the SHIL
signal on solution quality~\cite{oim_roychowdhury}. The typical
effects of the process, temperature, and voltage~(PVT) variations on
the SHIL signal, on the other hand, have not been explored. This is
especially important for scaling of CMOS OIM implementations that do
not eliminate oscillators~(e.g.\ digital emulation and iterative
hamiltonian computation, e.g.~\cite{roychowdhury_digital_emu,
  yilmaz_gpu, fpga_han, kulkarni_sachi, gpu_asic_ising}). Research is
advancing on methods such as problem embedding, interconnect sparsity,
coupling precision~(e.g.~\cite{3sat, memristor_coupling_shukla}) to
increase the utility and decrease the implementation cost of CMOS OIM
solutions. Scalability to a larger number of variables across larger
floorplans, for any of these CMOS OIM implementations, necessitates
PVT variation-tolerant SHIL signal generation and distribution.

\begin{figure}
  \centering \resizebox{0.85\columnwidth}{!}{ \input{./circuitikz_diagrams/stacked_ising.tex} }
  \caption{Illustration of an ROA brick delivering SHIL signals, an
    ROA-SHIL, stacked on top of an Oscillator-based Ising
    Machine~(OIM) layer}\label{fig:node_diag}
\end{figure}
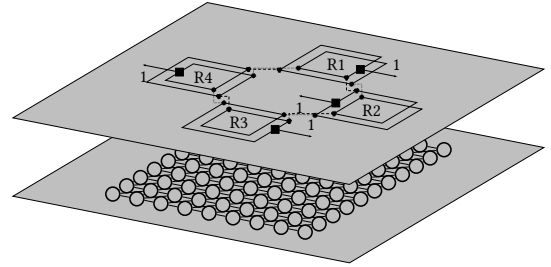

This paper introduces ROA-SHIL, using RTWOs organized into ROA bricks
as a centralized, variation-tolerant SHIL driver for CMOS OIMs. The
ROA brick is much larger in area footprint than the individual Ising
nodes while acting as a SHIL signal generation and distribution
network at the same time. The use of an ROA-SHIL is similar to the use
of the ROA topology generation and distribution of RTWO clock
signals~(e.g.~\cite{roa_brick, roa_low_skew, rotasyn}). The ROA-SHIL
can be placed on a separate 3D IC layer as illustrated in
\figref{fig:node_diag}, in the interposer~\cite{roa_interposer} of a
2.5D IC, or on the same plane of an IC\@. The nodes marked with
marker~1 in \figref{fig:node_diag} generate signals with the same
phase on phase-locked RTWO blocks R1, R2, R3, R4 forming the ROA\@.
That is, the same phase high frequency SHIL signal is available at
various distributed points across the floorplan. The RTWOs inherently
generate high frequencies at low power~\cite{roa_brick}, ROA brick
topology planning of RTWOs~\cite{roa_brick} enables the ROA
brick-based SHILs~(ROA-SHILs) to lock in phase, increasing tolerance
to PVT variations~\cite{rotary_stability}.

\section{Technical Background}\label{sec:tech_background}
In \secref{sec:ising_machines}, the preliminaries of Ising machines
and SHILs are introduced. In \secref{sec:rosc_ising}, CMOS ring
oscillator-based OIM~\cite{kim_1968_nodes}, which serves as the OIM
target architecture in this paper, is reviewed. In
\secref{sec:rotary}, the ROA brick topology of RTWOs~\cite{roa_brick}
is reviewed.

\subsection{Ising Machines and SHILs}\label{sec:ising_machines}
Ising machines~\cite{ernst_ising} minimize the energy in the system by
using devices that naturally tend to decrease entropy when coupled
together. As derived in~\cite{oim_roychowdhury}, a simplified version
of the Ising Hamiltonian equation that describes the energy of a
system is $H(s) = -\sum_{i<j}{J_{ij}s_{i}s_{j}}$. The coupling between
adjacent nodes is described by $J_{ij}$. The node states, $s_i$
and $s_j$, are assigned to groups, these groups being -1 or 1 for
Ising machines, rounded to the nearest groups if the nodes do not
binarize into two distinct groups. Ideally, the Ising Hamiltonian
decreases over time and settles into the lowest energy. In hardware
implementations, the Ising Hamiltonian is at risk of not binarizing
into two distinct groups. Subharmonic injection locking~(SHIL) of OIMs
modeled by the modified Kuramoto
equation~\cite{unconvential_computation} shown in~\eqref{eq:kuramoto}
alleviates this problem by employing an external perturbation into the
system of nodes that binarizes the nodes into discrete groups.

\begin{equation}
  \label{eq:kuramoto} \frac{1}{f_1}\frac{d}{dt}\Delta\phi_i(t)=
  \sum_{\mathclap{j=1,j\neq i}}^{n}{J_{ij}F_{c}(\Delta\phi_i(t)-\Delta\phi_j(t))+F_{s}(2\Delta\phi_i(t))}
\end{equation}

\begin{figure}[!h]
  \centering
  \includegraphics[width=0.85\columnwidth]{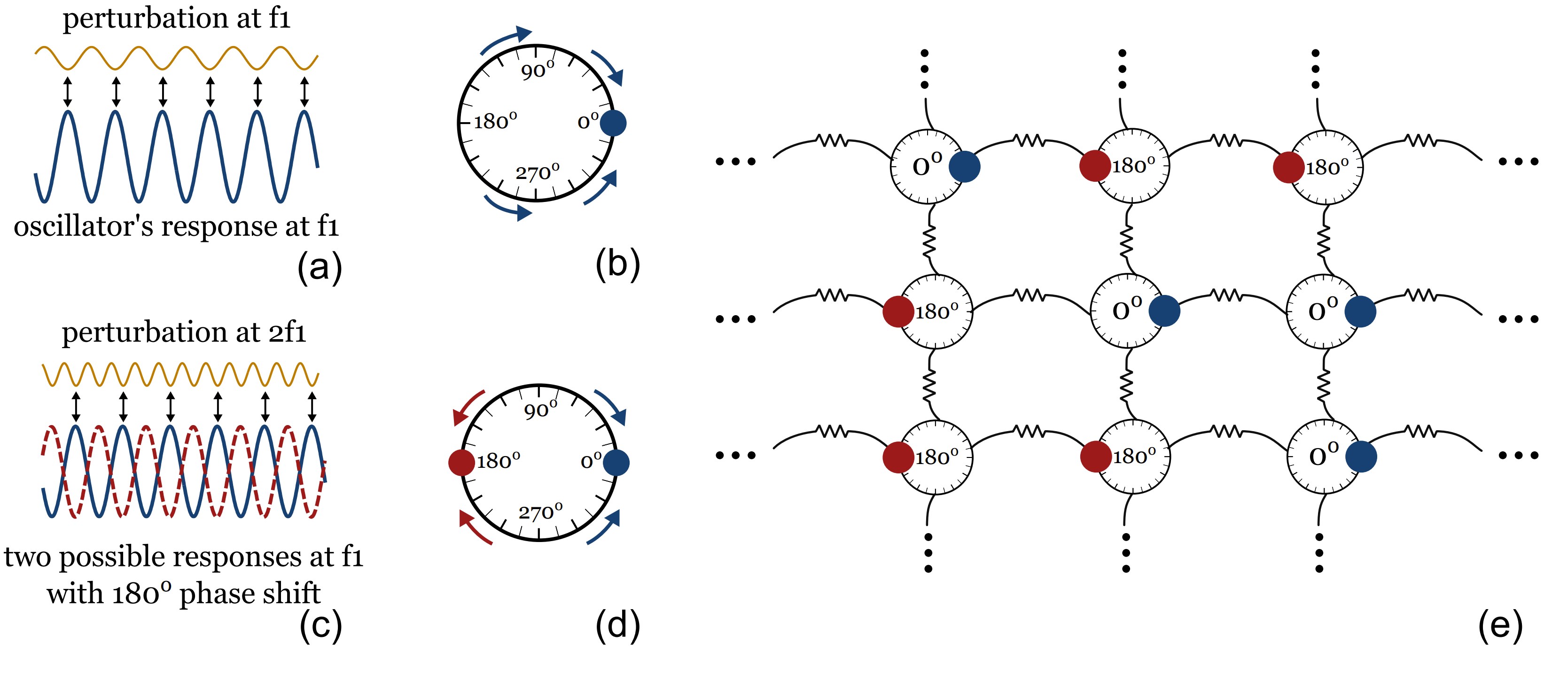}
  \caption{Injection locking mechanisms in oscillators: (a) oscillator
    at frequency is shifted due to perturbation frequency $f_1$; (b)
    oscillator locks to an arbitrary phase; (c) oscillator is bistable
    due to a perturbation of $2f_1$; (d) oscillator locks to one of
    two phases; (e) a system of oscillators being shifted to different
    phases due to subharmonic injection
    locking~\cite{oim_roychowdhury}}\label{fig:oim_shil}
\end{figure}

This equation shows a system of
oscillators interacting, while accounting for external perturbation
signals on each node. In~\eqref{eq:kuramoto}, $F_c(\cdot)$ captures
the phase change in coupled oscillators, denoted by $\Delta\phi_i(t)$
and $\Delta\phi_j(t)$, respectively, with $f_1$ denoting the nominal
frequency of the oscillator nodes. $F_s(\cdot)$ captures the impact of
2-SHIL as the external perturbation that binarizes phases as shown in
\figref{fig:oim_shil}~(d), in which a perturbation of $2f_1$ is injected
into each of the oscillators with the nominal frequency, $f_1$.

\subsection{ROSC Nodes for ROSC Ising Machine}\label{sec:rosc_ising}
\begin{figure}
  \centering\resizebox{0.75\columnwidth}{!}{
    \input{./circuitikz_diagrams/coupled_rosc.tex} }
  \caption{2-Node ROSC OIM with SHIL and parallel transmission gate
    coupling}\label{fig:rosc_connections}
\end{figure}
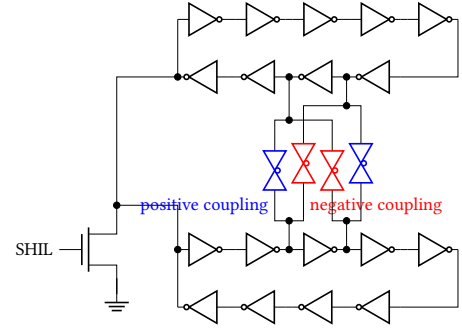

One implementation of oscillator-based Ising machines in
literature~\cite{kim_1968_nodes, oim_roychowdhury} employs ring
oscillators~(ROSCs) as the nodes and parallel transmission gates as
the coupling medium. Other implementations of CMOS~OIM Ising nodes and
coupling elements are possible, for
instance,~\cite{high_speed_phase_based} explores the use of RTWOs as
Ising nodes and~\cite{memristor_coupling_shukla} explores the use of
memristors as coupling elements for resolution. In this paper, the
implementation of an ROSC Ising machine with transmission gate-based
coupling elements~(called ROSC OIM and illustrated in
\figref{fig:rosc_connections}) is selected on which to introduce the
proposed ROA SHIL\@. For conformity to~\cite{kim_1968_nodes}, the ROSC
nodes are set up in a king's graph topography, a graph where the nodes
are connected in its cardinal and intercardinal directions to
neighboring nodes. As shown in \figref{fig:rosc_connections}, the
oscillators are either cross coupled for a negative edge or coupled at
a similar node for a positive edge, weight resolution $J_{ij}$ being
moderated through the number of parallel transmission gates. Mapping
of combinatorial optimization problems on Ising topologies~(NP-Hard
problems on Ising~\cite{ising_formulations}, max-cut~\cite{3sat,
  yilmaz_coupled_potts, multistage_potts, kim_1968_nodes, high_speed_phase_based},
MIMO~\cite{ising_mimo},~etc.), sparsity of Ising node
connections~(king's graph~\cite{kim_1968_nodes,
  high_speed_phase_based}, chimera graph~\cite{dwave},
all-to-all~\cite{kim_all_to_all}, sparsification~\cite{sparse_ising}),
and problem partitioning~\cite{huang_partitioning, fpga_partition},
and coupling programmability~\cite{memristor_coupling_shukla,
  fpga_han, yilmaz_gpu, roychowdhury_digital_emu} are some of the
ongoing directions of research.

\subsection{ROA Bricks}\label{sec:rotary}
\begin{figure}
  \centering
  \includegraphics[width=0.85\columnwidth]{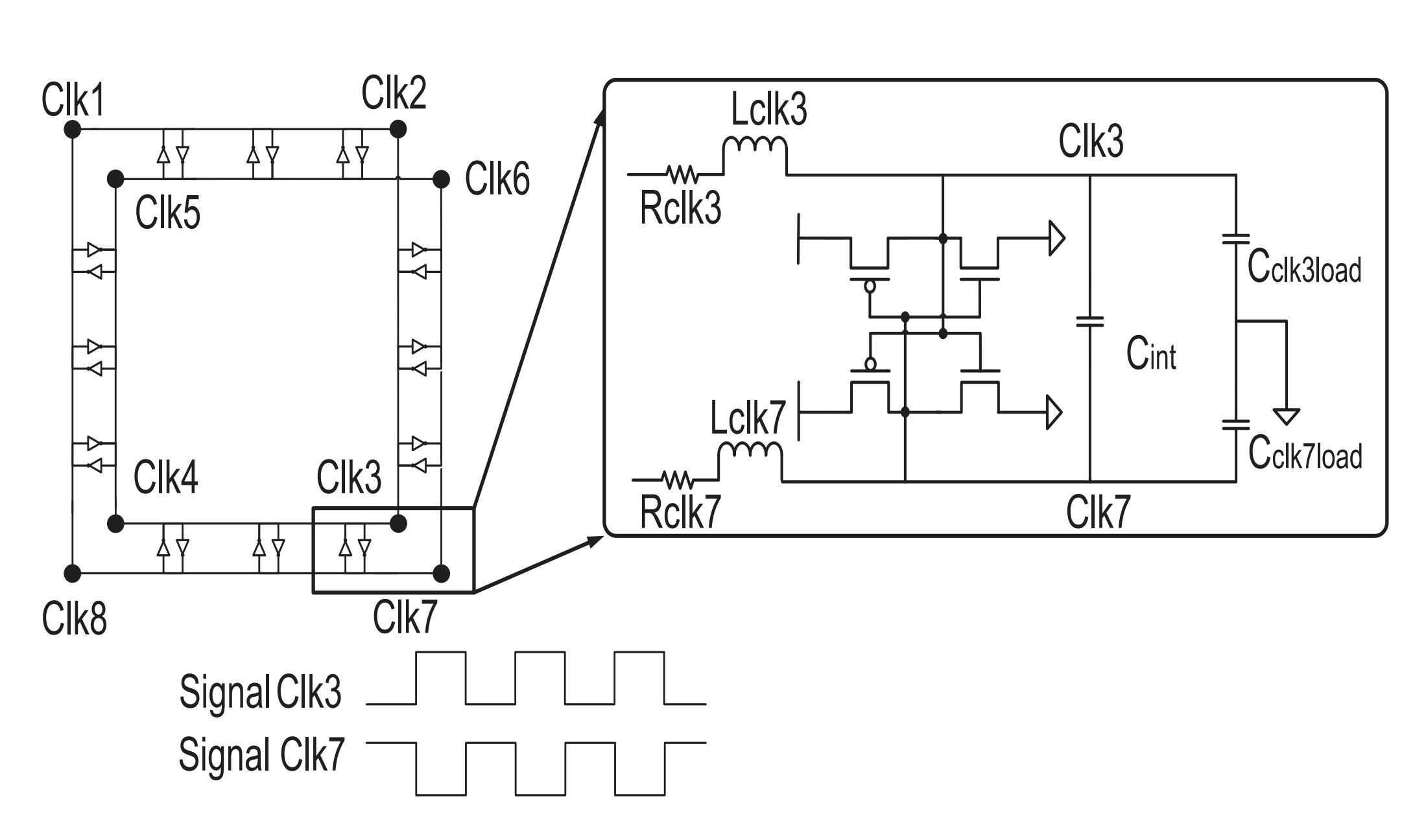}
  \caption{RTWO circuit diagram demonstrating mobius connected
    transmission lines, inverter pairs, and generated clock
    signals~\cite{low_freq_rtwo}. Each ring in \figref{fig:node_diag}
    is an RTWO circuit connecting in an ROA
    brick}\label{fig:rtwo_circ}
\end{figure}

Rotary Oscillatory Array~(ROA)~\cite{roa_brick} is a checkerboard
pattern of rotary traveling traveling wave oscillators connected
together. Rotary traveling wave oscillators~(RTWOs)~\cite{wood_rtwo}
employ metal lines twisted into a mobius pattern and connected with
cross-coupled inverters as displayed in \figref{fig:rtwo_circ}. The
inherent parasitics of the metal lines generate a resonant square wave
to travel across the rotary connected~(in a mobius topology) metal
line. RTWOs boast high frequency operation~(frequency of
\qty{3.2}{\GHz} to \qty{15}{\GHz} in~\cite{intel_rtwo}), and
demonstrate inversely proportionate power dissipation with increasing
frequency~\cite{rotasyn}.

Different topologies of RTWOs have been proposed in literature for
various design targets. For instance, zero clock skew~\cite{zeroa},
custom ASICs~~\cite{croa, sroa, rotasyn}, and 2.5D interposer
integration~\cite{ragh_mds}. The ROA brick topology~\cite{roa_brick} allows for the
waveform across multiple RTWO rings to be locked in phase at discrete
tapping points, used in~\cite{roa_brick} as a clock generation and
distribution network. RTWOs distributed using the OA brick
topology~(i.e.\ as a clock source) are shown
in~\cite{roa_load_balance} to demonstrate good tolerance to imbalance
in load capacitance due to placement and routing
optimizations. Another favorable feature of ROA bricks is the
resistance to variations: RTWOs are affected only by 2\% in frequency
and 4\% in phase in presence of a pessimistic +/- 10\% variation in
common PVT parameters~\cite{rotary_stability}.


\section{ROA Brick-Based Subharmonic Injection
  Locking~(ROA-SHIL)}\label{sec:scalable_shil}
The proposed use of an ROA brick of RTWO oscillators as the SHIL unit,
called ROA-SHIL, is illustrated in \figref{fig:node_diag}. Tapping
from the same phase points, marked~1 on \figref{fig:node_diag}, ROA
brick provides same phase, same frequency signals across the
floorplan. Similar to ROA bricks for clock signals
in~\cite{roa_brick}, proposed use of ROA bricks serves both as a
generation source and the distribution network for SHIL signals to OIM
nodes.

Two primary features of the proposed ROA-SHIL are the driving strength
and variation tolerance. The driving strength is high due to i)~the
injection point from the ROA bricks, marked 1 in
\figref{fig:node_diag}, acting as the driver, and ii)~having multiple
drivers thanks to same phase points. The variation tolerance is higher
thanks to the locked-in phase property of RTWOs on the ROA brick,
where impacts of variation are counter-balanced by the entire resonant
system~\cite{rotary_stability}. To this author's knowledge, there are
no on-chip integrated SHILs in literature. In~\cite{kim_1968_nodes},
an external perturbation source is employed to induce SHIL\@. The
distribution of the external SHIL signal to Ising nodes has not been
discussed.

\section{Analysis of ROA-SHIL}
To evaluate performance, three Ising implementations are compared. All
three OIM implementations are similar to~\cite{kim_1968_nodes} in
having ROSC-based Ising nodes~(i.e.\ ROSC OIMs) connected in the
king's graph topology with transmission gate based coupling between
nodes. The implementations differ in their SHIL source:
\begin{enumerate}
\item Ideal SHIL, i.e.\ from an external source such as
  in~\cite{kim_1968_nodes}, with a zero-delay model interconnect,
\item Distributed ring oscillator-based SHIL~(ROSC-SHIL)
  conceptualized in this paper as an alternative SHIL implementation
  for comparison purposes,
\item Proposed ROA brick-based SHIL~(ROA-SHIL).
\end{enumerate}
The experimental setup is described in \secref{sec:experiment}. The
driving strength and area utilization profiles are explored in
\secref{sec:drive_strength}. The results of these experiments are
utilized in \secref{sec:results} in choosing the number of Ising nodes
driven by each SHIL\@. The variation tolerance is investigated in
\secref{sec:variations}.

\subsection{Experimental Setup}\label{sec:experiment}
For hardware performance analysis, OIMs with ROSC-based Ising nodes
and transmission gate-based coupling with i)~an ideal SHIL, ii)~a
ROSC-SHIL, iii)~an ROA-SHIL are physically designed in a commercial
65nm PDK\@. Parasitic extraction for the ring oscillators~(i.e.~of the
ROSC OIM Ising nodes and the ROSC-SHIL) is performed using Siemens
Calibre PEX\@. The FEM modeling for the ROA-SHIL, using the process
in~\cite{rtwo_modeling}, is performed using Cadence EMX\@.

\begin{table}
  \caption{Ideal, ROSC-SHIL, and ROA-SHIL solving a 324-node ROSC OIM
    max-cut problem under nominal PVT
    conditions}\label{tab:shil_comparison}
  \begin{minipage}{\columnwidth}
  \centering
  \begin{tabular}{lccc}
    \toprule
    & ideal SHIL & ROSC-SHIL & ROA-SHIL \\
    \midrule
    OIM Structure & \multicolumn{3}{c}{324 node ROSC OIM operating at
                    1.13GHz} \\
    SHIL Structure & input pin & 5 ROSCs & 4 RTWOs \\
    SHIL Frequency & \qty{2.26}{\GHz} & \qty{2.18}{\GHz} & \qty{2.31}{\GHz} \\
    Total Power & \qty{458.9}{\mW} & \qty{468.2}{\mW} & \qty{502.0}{\mW} \\
    Total SHIL Power & \qty{0}{\mW}\footnote{Requires a distribution
                       network from pin to Ising nodes} & \qty{6.0}{\mW} & \qty{33.0}{\mW} \\
    Total Node Power & \qty{458.9}{\mW} & \qty{462.2}{\mW} & \qty{469.0}{\mW} \\
    Energy to Solution & \qty{8.71}{\nano\joule} & \qty{11.24}{\nano\joule} & \qty{11.20}{\nano\joule} \\
    Cycles to Solution & 23 cycles & 28 cycles & 26 cycles \\
    Solution Time & \qty{19.63}{\ns} & \qty{24.70}{\ns} & \qty{22.96}{\ns} \\
    Solution Accuracy & 95.2\% & 94.3\% & 94.6\% \\
    \bottomrule
  \end{tabular}
  \end{minipage}
\end{table}

The ROSC OIM nodes under test simulate at a fundamental
frequency~($f_1$) of \qty{1.13}{\GHz}. The ideal SHIL signal is set to
the second-order harmonic ($2f_1$) of \qty{2.26}{\GHz}. The ROSC-SHIL
operates at \qty{2.18}{\GHz}, empirically designed to the target
frequency, given a reasonable design effort. Similarly, the proposed
ROA-SHIL operates at \qty{2.31}{\GHz}. Specifically, the ROA brick
maintains a master frequency of \qty{20.82}{GHz}, which is
subsequently divided by a factor of 9 to produce the \qty{2.31}{\GHz}
SHIL signal, the power of the TSPC frequency
dividers~\cite{tspc_freq_div} are included in the ROA-SHIL power
reported in \tabref{tab:shil_comparison}.

\subsection{Driving Strength and Area Utilization}\label{sec:drive_strength}
Before analyzing the accuracy and variation tolerance with the
proposed ROA-SHIL, an exploratory study is performed to determine the
driving strength. This is necessary because the variation analysis
takes into account the distribution of the SHIL nodes and the
interconnects on the floorplan. The ROA-SHIL is a generation and
distribution network, with the footprint of \qty{425811}{\um\squared},
as designed in this analysis, and can host a large number of Ising
nodes. In comparison, note that a ROSC-SHIL, conceptualized in this
paper as an alternative on-chip SHIL, has a footprint of
\qty{112}{\um\squared}, needing to be instantiated and distributed
across a larger floorplan.

\begin{table}
  \caption{Driving strength and footprint of an ROA-SHIL vs an
    ROSC-SHIL}\label{tab:drive_strength}
  \centering
  \begin{tabular}{lccc}
    \toprule
    & ROSC-SHIL & ROA-SHIL \\
    \midrule
    SHIL nodes & 1 & 4~($4\times$) \\
    Footprint & \qty{113}{\um\squared} & \qty{425812}{\um\squared}~($\qty{3768}{}\times$) \\
    ROSC Ising nodes per SHIL & 65 & \qty{2792}{}~($43\times$) \\
    \bottomrule
  \end{tabular}
\end{table}

The power dissipation of the ideal~(i.e.\ external) SHIL is
considered zero, neglecting the costs of off-chip delivery to on-chip
oscillator nodes. The driving strength of a \qty{2.18}{\GHz} ROSC-SHIL
and the proposed \qty{2.31}{\GHz} ROA-SHIL, both custom designed in
the 65nm technology, are tested with capacitive loads, observing the
waveform quality. In the target technology, one ROSC-SHIL can
drive $\approx 65$ ROSC-based Ising nodes, ignoring interconnect
load. The ROA-SHIL can drive $\approx \qty{2792}{}$ ROSC-based Ising
nodes. The driving capability of the ROA brick is distributed over the
4 SHIL nodes on the brick, around a floorplan of \qty{664.5}{\um} by
\qty{640.8}{\um}. For the ROSC-SHIL, on the other hand, a new
ROSC-SHIL is placed for every 65 Ising nodes, leading to a distributed
placement of ROSC-SHILs. As reported in~\tabref{tab:drive_strength},
the driving strength of ROA-SHIL is approximately $43 \times$ higher
than that of a~(low-cost) ROSC SHIL\@.


Given the high driving strength of the ROA-SHIL, another analysis is
performed exploring if an ROA-SHIL would still be the dominant factor
of floorplan size with the maximum amount of nodes an ROA-SHIL can
drive. For this analysis, the area utilization of the ROA-SHIL is
investigated on the footprint of the ROA-SHIL based OIM. While the ROA-SHIL has a large
footprint, the ROA-SHIL innately has unutilized area~(i.e.~on a 2D IC
implementation) which other blocks are able to occupy. For a 2.5D and
3D IC implementation, as illustrated in \figref{fig:node_diag}, this
analysis provides insight to the footprint of the ROA-SHIL on one
layer compared to the footprint~(i.e.\ bounding box) of the Ising
nodes on another layer. Matching these footprints is useful~i)~in
having SHIL sources distributed in close proximity of the Ising nodes,
and~ii)~conceptualizing the use of the ROA-SHIL with co-placed Ising
nodes as a repeatable tile/IP block for OIMs much larger in size.


\begin{figure}
  \centering
  \includegraphics[width=0.85\columnwidth]{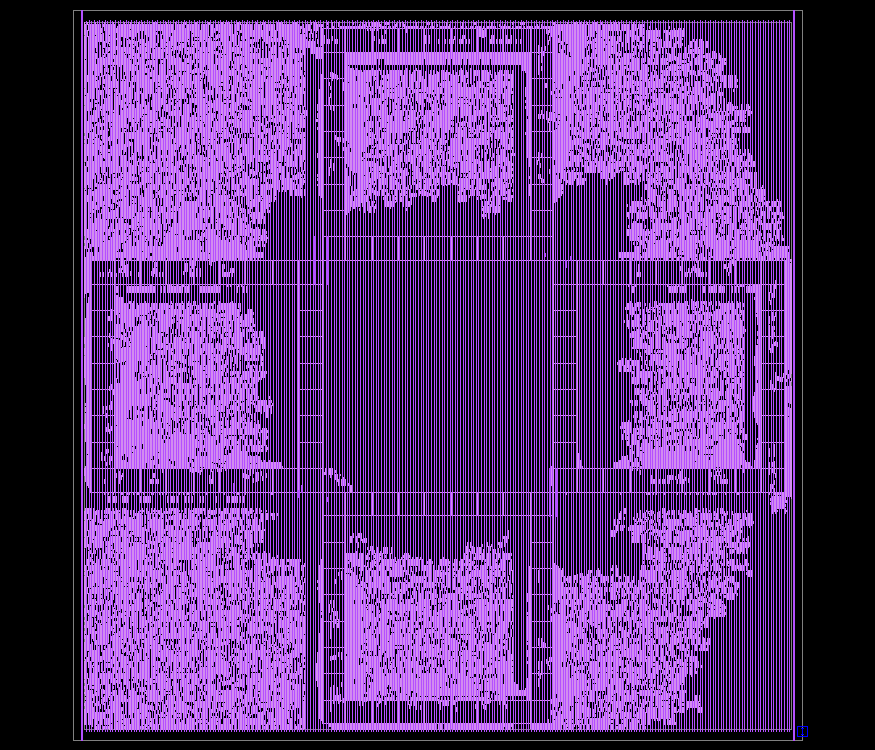}
  \caption{Placement of a $30 \times 30$ OIM with 900 ROSC ising nodes
    and 1 ROA-SHIL. ROA-SHIL serves as the SHIL signal generation
    source and provides global distribution}\label{fig:roa_layout}
\end{figure}

\begin{table*}[t]
  \caption{Variation analysis of proposed ROA-SHIL and ROSC-SHIL on
    ROSC OIM solving 324-node max-cut problem}\label{tab:variations}
  \centering
  \begin{tabular}{ll|c|cc|cc|c}
    \toprule
    & & Nominal & \multicolumn{2}{c|}{Voltage} & \multicolumn{2}{c|}{Temperature} & Process \\
    & & & 5\% & 10\% & -40\textdegree C & 125\textdegree C & \\
    \midrule
    \multirow{2}{*}{ROSC-SHIL} & Binarized Nodes & Yes & No & No & Yes & Yes & No \\
                 & Solution Accuracy & 94.3\% & 78.4\% & 75.7\% & 95.0\% & 93.3\% & 90.0{-}95.0\% \\
    \midrule
    \multirow{2}{*}{ROA-SHIL} & Binarized Nodes & Yes & Yes & Yes & Yes & Yes & Yes \\
                 & Solution Accuracy & 94.6\% & 96.9\% & 96.0\% & 96.1\% & 96.3\% & 93.0{-}95.0\% \\
    \bottomrule
  \end{tabular}
\end{table*}

\begin{figure}
  \centering
  \includegraphics[width=0.85\columnwidth]{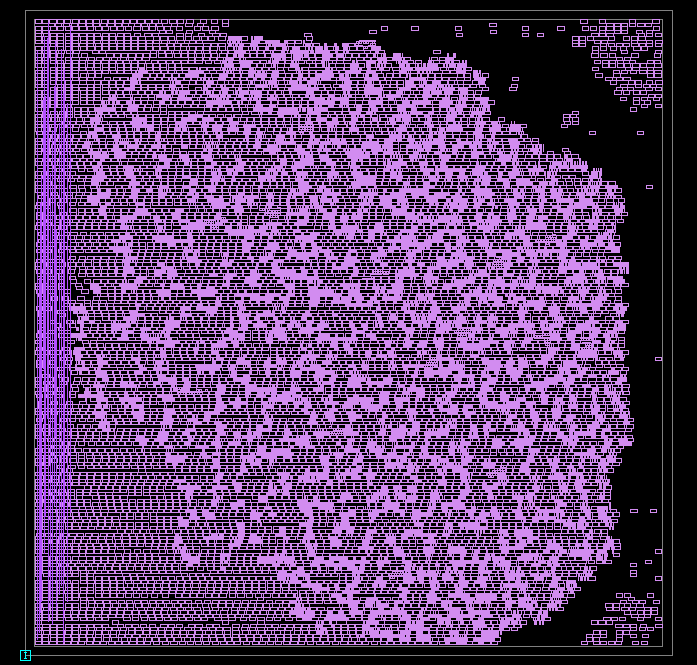}
  \caption{Placement of a $30 \times 30$ OIM with 900 ROSC nodes and 14 (distributed) ROSC-SHILs}\label{fig:rosc_layout}
\end{figure}

Layouts with different number of Ising nodes are designed with an
arbitrarily selected floorplan utilization factor
of $\approx 60\%$. It is observed, empirically, that as the OIM graph
size exceeds 625 nodes, a size that is less than the
\qty{2792}{}/4=698 node driving strength of one tapping point in an ROA brick,
the total footprint becomes dominated by the individual
Ising nodes rather than the ROA-SHIL\@. For graph sizes below this
threshold, the ROA-SHIL increases the overall design footprint,
resulting in a floorplan utilization below 60\%. The majority of
combinatorial optimization problems of interest exceed this number of
Ising nodes~\cite{kim_1968_nodes, high_speed_phase_based}. At
equivalent node counts of 625 ROSC Ising nodes, floorplans for both
ROSC-SHIL and proposed ROA-SHIL exhibit comparable
dimensions: The ROA-SHIL layout~\figref{fig:roa_layout}
measures \qty{692.3}{\um} in width, while the ROSC-SHIL
layout~\figref{fig:rosc_layout} measures \qty{689.4}{\um}. As the
system scales, the ROSC-SHIL architecture requires the
integration and synchronization of additional units, thereby
increasing the area overhead. Conversely, the ROA-SHIL
maintains the ability to drive a significantly larger quantity of
nodes~(i.e.~$\approx \qty{2792}{}$) within a single structure.

\subsection{324-Node OIM Performance}\label{sec:results}
All three SHIL implementations under analysis are configured to solve
the same, arbitrarily selected max-cut problem consisting of 324 nodes
arranged in an $18\times18$ graph. The resulting cut sizes are
benchmarked for accuracy against a software-based baseline using
D-Wave’s Tabu solver~\cite{qubo_tut}. \tabref{tab:shil_comparison}
presents a comparative analysis of the three evaluated SHIL
architectures. The baseline for comparison is the ideal SHIL
operating at \qty{2.26}{\GHz} (the exact second harmonic of the
\qty{1.13}{\GHz} node frequency with ideal interconnects). To
synchronize the 324 OIM nodes, the baseline architecture utilizes five ROSC-SHIL units driving the
maximum amount of nodes without degrading the waveform, $\approx 65 \times 5 = 325$ Ising nodes. Given the slight
deviation in operating frequencies, total power consumption is evaluated relative to
frequency. The overhead of power dissipated by the SHIL is reported
for each respective type. For the 324-node OIM configuration, a single
ROA brick comprising four RTWO rings is employed, remaining under the
driving capacity.

The baseline architecture of an ideal SHIL on an OIM exhibits a power
consumption of \qty{458.9}{\mW}. Due to the power dissipation of the
ROA-SHIL being much higher compared to that of an
ROSC-SHIL~(\qty{33.0}{\mW} vs \qty{6.0}{\mW}), the total power
dissipation of the OIM with the proposed ROA-SHIL is $\approx 7.2\%$
higher~(\qty{502}{\mW} vs \qty{468.2}{\mW}). However, the
energy-to-solution with the proposed ROA-SHIL induced OIM is
lower~(\qty{11.20}{\nano\joule} vs \qty{11.24}{\nano\joule}) than the
energy-to-solution of the ROSC-SHIL induced OIM\@. This
energy-to-solution advantage is due to a slightly faster solution time
with the proposed ROA-SHIL induced OIM, a metric that improves with a
faster OIM node and SHIL
frequency~\cite{high_speed_phase_based}. Under nominal operating
conditions, the accuracy across all three implementations remains
consistent~(95.2\% for an ideal SHIL, 94.3\% for a ROSC SHIL, and
94.6\% for an ROA-SHIL).

\subsection{Effects of Process Variations on Accuracy}\label{sec:variations}
The ROA-SHIL and ROSC-SHIL OIM architectures are characterized
under process, temperature, and voltage~(PVT) variations. In these
analyses, the supply voltage is deviated by $\pm$5\% and $\pm$10\%, while
temperature is evaluated at the corners of -40°C and
125°C. Process variations are modeled using a Gaussian distribution a
specified by the PDK\@. These studies utilize the 324-node max-cut
problem established in \secref{sec:results}.

Results of the variation analysis are summarized in
\tabref{tab:variations}. Using the proposed ROA-SHIL, OIM solution
accuracy increases by approximately 2.4\%~(reaching 97\% from a
nominal 94.6\%) in the presence of a 5\% supply voltage variation. At
a 10\% voltage deviation, accuracy decreases slightly relative to the
accuracy of the OIM with a 5\% voltage variation but remains elevated
above the nominal voltage baseline at 96\%. This performance demonstrates
significant tolerance to voltage fluctuations and aligns with the
findings in~\cite{oim_roychowdhury}, where fluctuating SHIL signals
are shown to enhance solution accuracy. Due to their inherent
resistance to variations, ROA bricks maintain stable SHIL frequency
and phase, ensuring consistent OIM performance across the tested
voltage range.

Conversely, supply voltage fluctuations prove catastrophic for OIMs
induced with ROSC-SHILs, as detailed in \tabref{tab:variations}. Under
these variations, the OIM does not exhibit effects of injection
locking. The ROSC OIM with the ROSC-SHIL degrades solution accuracy to
78.4\% at a 5\% voltage variation and further declines to 75.7\% at a
10\% voltage variation. These significant reductions in accuracy are
primarily attributed to the high sensitivity of ring oscillator
frequency and phase to supply voltage shifts. This is different than
the intentional fluctuating the magnitude of the
SHIL~(i.e.~\cite{oim_roychowdhury}) as the PVT variations are
unavoidable, and will prevent intended values of fluctuations, leading
to failures in locking into 2-SHIL stability regions.

Thermal variations do not significantly impact the accuracy of either
the ROSC-SHIL or the proposed ROA-SHIL induced OIMs, with deviations
remaining within a $\pm$1\% margin. As shown in
\tabref{tab:variations}, the accuracy of proposed ROA-SHIL induced
OIMs are 96.1\% and 96.3\% at the two temperature corners
-40\textdegree C and 125\textdegree C, respectively. The accuracy at
both corners are higher than the 94.6\% accuracy at the nominal
corner.

Process variations result in moderate accuracy degradation but lead to
failures in locking into 2-SHIL stability regions. Across Monte Carlo
iterations, the ROA-SHIL induced OIMs maintain accuracy within
1\%~(93–95\%) of the nominal results, whereas for the ROSC-SHIL induced OIMs, accuracy
decreases to 90\%. While these accuracy drops appear marginal in some
contexts, the ROSC-based OIM nodes fail to binarize in any of the
process-variant simulation runs, unlike the ROA-SHIL which ensures
successful binarization across all iterations.

\section{Conclusion}\label{sec:conclusion}
This work proposes the utilization of rotary oscillator array~(ROA)
based RTWO structures for subharmonic injection locking~(SHIL). This
ROA-SHIL architecture facilitates the generation and distribution of
external perturbation signals with high tolerance to process, voltage,
and temperature~(PVT) variations. Experimental results demonstrate
that ROA-SHILs exhibit superior robustness compared to~(not used in
literature, but conceptualized in this work) ROSC-SHIL
implementations. Specifically, under adverse PVT conditions,
ROSC-SHILs fail to achieve phase binarization, and lead to significant
degradation in solution quality. Conversely, the proposed ROA-SHILs
maintain performance parity with ideal, off-chip SHIL signals,
ensuring high solution accuracy and reliable system operation across
varying process and environmental conditions.

\clearpage

\bibliographystyle{ACM-Reference-Format}
\bibliography{ref}

\end{document}

%% file: circuitikz_diagrams/stacked_ising.tex
\tdplotsetmaincoords{70}{120}
\begin{tikzpicture}[tdplot_main_coords]

\newcommand{\layer}[4]{%
  \begin{scope}[canvas is xy plane at z=#1]

      \draw[fill=lightgray, opacity=1.0] (-7,-7) rectangle (7,7);
    \node(new_node) at #4 {#3};
  \end{scope}
}


\layer{-2}{gray}{
  \begin{tikzpicture}[tdplot_main_coords]
    \input{./circuitikz_diagrams/big_node_diagram.tex}
  \end{tikzpicture}
}{(0,0)}

\layer{1}{gray}{
  \begin{tikzpicture}[tdplot_main_coords,
    scale=0.5,
    every node/.style={scale=0.5},
  ]
    \input{./circuitikz_diagrams/roa_brick.tex}
  \end{tikzpicture}
}{(0,0)}

\end{tikzpicture}

%% file: circuitikz_diagrams/coupled_rosc.tex
\usetikzlibrary{calc}

\newsavebox{\shil}
\sbox{\shil}{
  \begin{tikzpicture}[scale = 1, transform shape, color=red]
	\ctikzset{
        logic ports=ieee,
        logic ports/scale=0.25,
        logic ports/fill=white
	}

	\draw node(inv1)[not port] at (1, 1) {};
	\draw node(inv2)[not port] at (2, 1) {};
	\draw node(inv3)[not port] at (3, 1) {};
	\draw node(inv4)[not port] at (4, 1) {};
	\draw node(inv5)[not port] at (5, 1) {};
	\draw node(inv9)[not port, xscale=-1] at (1, 0) {};
	\draw node(inv8)[not port, xscale=-1] at (2, 0) {};
	\draw node(inv7)[not port, xscale=-1] at (3, 0) {};
	\draw node(inv6)[not port, xscale=-1] at (4, 0) {};
    \draw (inv1.out) -- (inv2.in);
    \draw (inv2.out) -- (inv3.in);
    \draw (inv3.out) -- (inv4.in);
    \draw (inv4.out) -- (inv5.in);
    \draw (inv5.out) |- (inv6.in);
    \draw (inv6.out) -- (inv7.in);
    \draw (inv7.out) -- (inv8.in);
    \draw (inv8.out) -- (inv9.in);
    \draw (inv9.out) -- (inv1.in);

  \end{tikzpicture}
}

\tikzset{
    pics/rosc/.style={
      code={

        \draw node(inv1)[not port] at (1, 1) {};
        \draw node(inv2)[not port] at (2, 1) {};
        \draw node(inv3)[not port] at (3, 1) {};
        \draw node(inv4)[not port] at (4, 1) {};
        \draw node(inv5)[not port] at (5, 1) {};
        \draw node(inv9)[not port, xscale=-1] at (1, 0) {};
        \draw node(inv8)[not port, xscale=-1] at (2, 0) {};
        \draw node(inv7)[not port, xscale=-1] at (3, 0) {};
        \draw node(inv6)[not port, xscale=-1] at (4, 0) {};
        \draw (inv1.out) -- (inv2.in);
        \draw (inv2.out) -- (inv3.in);
        \draw (inv3.out) -- (inv4.in);
        \draw (inv4.out) -- (inv5.in);
        \draw (inv5.out) |- (inv6.in);
        \draw (inv6.out) -- (inv7.in);
        \draw (inv7.out) -- (inv8.in);
        \draw (inv8.out) -- (inv9.in);
        \draw (inv9.out) -- (inv1.in);
}}
}

\begin{tikzpicture}
  \ctikzset{
    logic ports=ieee,
    logic ports/scale=0.5,
    logic ports/fill=white
  }
  \node[nmos](shil_tran) at (0,1){};
  \draw (shil_tran) ++(0, -0.5) node [ground](shil_gnd){};
  \draw (shil_tran) ++(-1, 0) node[left] (shil_out){SHIL};

  \draw (shil_tran.source) -- (shil_gnd.center);
  \draw (shil_tran.gate) -- (shil_out);


  \pic(rosc1) at (0.5,4) {rosc};
  \pic(rosc2) at (0.5,0) {rosc};

  \draw (shil_tran.drain) node[circ]{} |- (rosc1inv9.out) node[circ]{};
  \draw (shil_tran.drain) -| (rosc2inv1.in) node[circ]{};

  \node[circ](rosc1_junc1) at ($ (rosc1inv8.in)!0.5!(rosc1inv7.out) $){};
  \draw (rosc1_junc1) -- ++(0, -0.75) node [circ](tran1_junc1){};
  \node[circ](rosc1_junc2) at ($ (rosc1inv7.in)!0.5!(rosc1inv6.out) $){};
  \draw (rosc1_junc2) -- ++(0, -0.5) node [circ](tran2_junc1){};
  \node[circ](rosc2_junc1) at ($ (rosc2inv3.in)!0.5!(rosc2inv2.out) $){};
  \draw (rosc2_junc1) -- ++(0, 0.5) node [circ](tran1_junc2){};
  \node[circ](rosc2_junc2) at ($ (rosc2inv4.in)!0.5!(rosc2inv3.out) $){};
  \draw (rosc2_junc2) -- ++(0, 0.5) node [circ](tran2_junc2){};

  \draw (tran1_junc1) -- ++(-0.25, 0) to[inline tgate, transform
  shape, color=blue] ($(tran1_junc2)+(-0.25,0)$) node[above left, blue] {positive coupling} -- (tran1_junc2);
  \draw (tran2_junc1) -- ++(-0.75, 0) to[inline tgate, transform
  shape, color=red] ($(tran1_junc2)+(0.25,0)$) node[above right, red] {negative coupling} -- (tran1_junc2)-- (tran1_junc2);
  \draw (tran1_junc1) -- ++(0.75, 0) to[inline tgate, transform
  shape, color=red] ($(tran2_junc2)+(-0.25,0)$) -- (tran2_junc2);
  \draw (tran2_junc1) -- ++(0.25, 0) to[inline tgate, transform
  shape, color=blue] ($(tran2_junc2)+(0.25,0)$) -- (tran2_junc2);


\end{tikzpicture}